\PassOptionsToPackage{table,xcdraw}{xcolor}

\documentclass[sigconf,natbib=true,authorversion]{acmart}

\usepackage[table,xcdraw]{xcolor}
\usepackage{multicol}
\usepackage{multirow}
\usepackage[normalem]{ulem}
\useunder{\uline}{\ul}{}
\usepackage{hyperref}
\usepackage{graphicx}
\usepackage{subfigure}
\usepackage{fixltx2e}
\usepackage[normalem]{ulem} 

\usepackage{adjustbox} 
\newcommand{\cmark}{\checkmark}%
\newcommand{\xmark}{$\times$}%
\usepackage{enumitem}

\usepackage{etoolbox}
\makeatletter
\apptocmd{\@mkauthors@iii}{%
  \global\setbox\mktitle@bx=\vbox{\noindent
    \box\mktitle@bx\par\vskip3pt\nointerlineskip
    {\centering\large
     BARS Benchmark: \textcolor{magenta}{\url{https://recbench.github.io/BARS}}\par}%
    \vskip12pt}%
}{}{\ClassWarning{CIKM2021}{Failed to patch \@mkauthors@iii}}
\makeatother

\copyrightyear{2022} 
\acmYear{2022} 
\setcopyright{acmlicensed}\acmConference[SIGIR '22]{Proceedings of the 45th International ACM SIGIR Conference on Research and Development in Information Retrieval}{July 11--15, 2022}{Madrid, Spain}
\acmBooktitle{Proceedings of the 45th International ACM SIGIR Conference on Research and Development in Information Retrieval (SIGIR '22), July 11--15, 2022, Madrid, Spain}
\acmPrice{15.00}
\acmDOI{10.1145/3477495.3531723}
\acmISBN{978-1-4503-8732-3/22/07}

\begin{document}

\title{BARS: Towards Open Benchmarking for Recommender Systems}
\fancyhead{} 

\author{Jieming Zhu$^{\mathsection}$}
\affiliation{%
  \institution{Huawei Noah's Ark Lab}
  \city{Shenzhen}
  \country{China}
}
\email{jiemingzhu@ieee.org}

\author{Quanyu Dai}
\affiliation{%
  \institution{Huawei Noah's Ark Lab}
  \city{Shenzhen}
  \country{China}
}

\author{Liangcai Su}
\affiliation{%
  \institution{Tsinghua University}
  \city{Shenzhen}
  \country{China}
}
\author{Rong Ma}
\affiliation{%
  \institution{Tsinghua University}
  \city{Beijing}
  \country{China}
}

\author{Jinyang Liu$^{\star}$}
\affiliation{%
  \institution{The Chinese University of Hong Kong}
  \city{Hong Kong}
  \country{China}
}

\author{Guohao Cai}
\affiliation{%
  \institution{Huawei Noah's Ark Lab}
  \city{Shenzhen}
  \country{China}
}

\author{Xi Xiao$^{\mathsection}$}
\affiliation{%
  \institution{Tsinghua University}
  \city{Shenzhen}
  \country{China}
}
\email{xiaox@sz.tsinghua.edu.cn}

\author{Rui Zhang}
\affiliation{\country{\url{ruizhang.info}}}
\email{rayteam@yeah.net}

\thanks{$^\star$ Part of the work was done when the author studied at Sun Yat-Sen University.}
\thanks{$^\mathsection$ Corresponding authors}

\renewcommand{\authors}{Jieming Zhu, Kelong Mao, Quanyu Dai, Liangcai Su, Rong Ma, Jinyang Liu, Guohao Cai, Zhicheng Dou, Xi Xiao, Rui Zhang}





\begin{abstract}
The past two decades have witnessed the rapid development of personalized recommendation techniques. Despite significant progress made in both research and practice of recommender systems, to date, there is a lack of a widely-recognized benchmarking standard in this field. Many existing studies perform model evaluations and comparisons in an ad-hoc manner, for example, by employing their own private data splits or using different experimental settings. Such conventions not only increase the difficulty in reproducing existing studies, but also lead to inconsistent experimental results among them. This largely limits the credibility and practical value of research results in this field. To tackle these issues, we present an initiative project (namely BARS) aiming for open benchmarking for recommender systems. In comparison to some earlier attempts towards this goal, we take a further step by setting up a standardized benchmarking pipeline for reproducible research, which integrates all the details about datasets, source code, hyper-parameter settings, running logs, and evaluation results. The benchmark is designed with comprehensiveness and sustainability in mind. It covers both retrieval and ranking tasks, and also enables researchers to easily follow and contribute to the research in this field. This project will not only reduce the redundant efforts of researchers to re-implement or re-run existing baselines, but also drive more solid and reproducible research on recommender systems. We encourage the community to adopt the BARS benchmark for future evaluations and welcome contributions to the project.








\end{abstract}



\maketitle

\section{Introduction}
Personalized recommendation plays an indispensable role in our daily life to help people discover information of their interests. It has been adopted in a wide range of online applications, such as E-commerce, social media, news feeds, and music streaming. The application of recommender systems not only helps users alleviate information overload but also serves as an important source for platform vendors to gain revenues from user interactions (e.g., clicks, downloads, or purchases). Typically, modern recommender systems in industry consist of two major phases, namely retrieval and ranking~\cite{YouTubeNet}. The retrieval phase (interchangeably referred to as the matching phase) aims for efficient retrieval of candidate items and helps reduce the candidate item set from millions to hundreds while guaranteeing a high recall rate. The ranking phase further ranks the retrieved items by leveraging abundant user/item/context features and returns the top-k (\textit{e.g.}, in tens) items to users. 

Under the wave of deep learning, we have witnessed the
rapid development of recommendation techniques in both phases. Many deep neural networks, such as auto-encoders~\cite{MVAE, MacridVAE, RecVAE},  graph neural networks (GNN)~\cite{GLP, LightGCN,NIA-GCN,FiGNN}, convolution neural networks~\cite{CNN-FeatureGen}, and transformers~\cite{BERT4Rec, AutoInt}, have been applied to recommender systems for better retrieval and ranking capabilities. Moreover, numerous new recommendation algorithms are published every year. Despite the widespread research of recommender systems, there is still a lack of a widely-recognized benchmarking standard in this field. This raises a growing concern that a significant number of research papers lack rigor in evaluation and reproducibility of reported results~\cite{Repsys13_Joseph,are_we_really,are_we_eval}. In the literature, existing studies on recommender systems are often evaluated in an ad-hoc way, for example, by employing their own private data splits, using a different experimental setting, or adopting a weak baseline (e.g., not well tuned). In many cases, the reported results cannot be easily reproduced due to the lack of either data preprocessing details, model implementations, hyper-parameter configurations, or even all of them. This not only leads to inconsistent results among existing studies but also makes fairly comparing model effectiveness a non-trivial task, which largely limits the credibility and practical value of research results in this field~\cite{Impact_recsys}. Recently, the critical issue has attracted more and more attention and has been reported in several pioneer studies~\cite{SachdevaM20, CIKM20_Critique, ncf_vs_mf, are_we_eval, are_we_really}.

Inspired by the success of the ImageNet benchmark~\cite{ImageNet} in the CV domain and the GLUE benchmark~\cite{GLUE} in the NLP domain, we propose an initiative for benchmarking recommender systems related research in a standard and reproducible manner. Many precious research efforts have been made towards this goal, which can be broadly categorized into three groups: \textit{\textbf{1) Benchmarking datasets}}. A lot of real-world benchmarking datasets have been released to promote research for recommender systems, including MovieLens~\cite{MovieLens}, Yelp~\cite{Yelp2018}, AmazonBooks~\cite{AmazonBooks_data}, Gowalla~\cite{Gowalla}, Criteo~\cite{Criteo}, just to name a few. Furthermore, a large set of preprocessed datasets are readily available in RecBole~\cite{RecBole}. \textit{\textbf{2) Benchmarking tools}}. With the evolution of deep learning-based recommendation techniques, many useful open-source libraries and tools have been developed, such as DeepRec~\cite{DeepRec}, daisyRec~\cite{are_we_eval}, NeuRec~\cite{NeuRec}, 
RecBole~\cite{RecBole},
DeepCTR~\cite{DeepCTR}, 
FuxiCTR~\cite{FuxiCTR_git}, 
TensorFlow Recommenders~\cite{tfrs}, TorchRec~\cite{TorchRec}, 
PaddleRec~\cite{paddlerec}, 
and EasyRec~\cite{EasyRec}, each providing tens of popular recommendation algorithms\footnote{It is worth mentioning that sometimes benchmarking datasets or tools are also referred to as ``benchmarks" in the literature, since they provide necessary data and components (e.g., data loader, metric evaluator, hyperparameter tuner, etc.) for benchmarking use. Yet, we target at a GLUE-like leaderboard as the benchmark in this paper, and refer to the process to obtain such a leaderboard as the benchmarking pipeline.}. \textit{\textbf{3) Benchmarking results}}. Some recent work performs evaluation studies to reproduce existing methods or identify their non-reproducibility issues in deep learning-based recommendation~\cite{are_we_eval}, session-based recommendation~\cite{Recsys_benchmark}, and review-based recommendation~\cite{SachdevaM20}. However, none of them provides a comprehensive leaderboard of benchmarking results with high reproducibility. In this work, we aim to chain the above pieces of work together (i.e., datasets, tools, and results) to establish BARS, a standardized open benchmarking pipeline for recommender systems.

Two closely-related studies to ours are Elliot~\cite{Elliot} and Recbole~\cite{RecBole}, which provide rigorous evaluation frameworks for developing and benchmarking recommendation models. In contrast to some earlier attempts~\cite{RecBench,benchmarking_rs}, the design of our benchmarking system has the following merits.



\begin{itemize}[leftmargin=*]
    \item \textbf{Standardization}: We propose a standardized benchmarking pipeline to promote reproducible research, which involves seven necessary artifacts. In addition to the widely-existing open datasets and open-source models, we also mark data splits, evaluation protocols, hyper-parameter configurations, running logs, evaluation results as required artifacts, which need to be properly recorded during benchmarking. The standardized pipeline not only encourages reuse of open datasets as well as open-source code, but also aims for tractable data splits and reproducible evaluation results. Without such a standard, existing studies may only deliver parts of the artifacts (e.g., only code in DeepCTR~\cite{DeepCTR}, lack of hyper-parameters and running logs in daisyRec~\cite{are_we_eval}).
    \item \textbf{Reproducibility}: Reproducibility is the first-class goal of our benchmark. For each result in BARS, we provide the corresponding reproducing steps in details (along with the required artifacts) to allow anyone for easy reproduction, including hardware and software environment, dataset and model configurations, as well as running scripts and logs. These could also serve as good guidelines for junior researchers.
    
    
    
   \item \textbf{Reusability}: On one hand, none of the datasets were created from scratch for the benchmark. We reuse pre-existing datasets (e.g., Criteo~\cite{Criteo}) because they have been implicitly agreed upon by the community due to their wide use in existing papers. We also reuse the open-source code wherever possible (e.g., daisyRec~\cite{are_we_eval}) to encourage the contributions from different parties. On the other hand, 
    our work aims to provide well-tuned benchmarking results and settings that could be directly reused for future research. Concretely, we evaluate the performance of tens of state-of-the-art (SOTA) models and provide strong baselines to compare with. Given our benchmarking results, researchers could easily gauge the effectiveness of new models, while largely reduce the tedious yet redundant efforts to re-implement and re-run the baselines for publishing a new paper. 
    \item \textbf{Comprehensiveness}: To aim for comprehensiveness of our BARS benchmark, we have integrated over 6 preprocessed datasets and more than 70 recommendation algorithms covering both retrieval and ranking phases. The thorough benchmarking results on these models and datasets provide a systematic view of the effectiveness of current recommendation algorithms and facilitate better understanding of the real progress that we have made. To the best of our knowledge, our leaderboard provides the most comprehensive benchmarking results in the recommender systems community.
    \item \textbf{Sustainability}: A benchmark will become stale soon if it is not regularly maintained and widely recognized by the community. To ensure sustainability of our benchmark, we build a benchmark website like GLUE~\cite{GLUE} that allows anyone in the community to easily follow and contribute. We call for contributions from any aspect to foster sustainable development of BARS, including extending new datasets, benchmarking existing models on new data splits, implementing new SOTA models, reporting improved results with better hyper-parameter settings, polishing the documentation, etc.
    \item \textbf{Industrial-level}: Existing studies~\cite{are_we_really,are_we_eval} focus primarily on conventional recommendation algorithms, such as collaborative filtering (CF), which fail to distinguish between retrieval and ranking tasks from an industrial view. Actually, CF is only one type of retrieval methods used in industrial recommender systems. Instead, our work explicitly classifies existing recommendation methods into retrieval and ranking phases, and performs benchmarking with different metrics on candidate item retrieval (CIR) and click-through rate (CTR) prediction tasks, respectively. Thus, our benchmark better fits the need for industrial use. 
    
 
\end{itemize}

Bearing these design goals in mind, we build the BARS benchmark. 
While the ultimate goal of BARS is to drive open benchmarking for all recommendation tasks, as an initiative work, our current version covers benchmarking results for candidate item retrieval and CTR prediction. In particular, we ran over 8,000 experiments for more than 15,000 GPU hours ($\sim$625 days) in a standard setting to benchmark more than 70 recommendation models on six widely-used datasets. The large number of experiments ensure sufficient hyper-parameter tuning of each model via grid search. Under sufficient model tuning, we found that the effectiveness of many recent methods may be exaggerated. Many of them have smaller differences than expected and sometimes are even inconsistent with what reported in the literature. We believe that easy access to consistently split and pre-processed datasets with reusable baseline results could help the research community avoid reporting inconsistent or misleading results in their future work. 


In summary, our work could serve as a useful resource for several different group of readers: \textit{{Researchers}} are easy to get a full view of model improvements made over recent years and also gauge the effectiveness of new models conveniently. \textit{{ Practitioners}} could reduce unnecessary trials by comparing different models via the benchmark leadboard. \textit{{Competitors}} can easily implement baseline solutions by leveraging our source code and  benchmarking scripts. \textit{{Educators}} could use BARS as a valuable resource to teach and study recommendation algorithms.


By setting up an open benchmarking leaderboard, together with the freely available benchmarking artifacts (e.g., datasets, code, configurations, results, and reproducing steps), we hope that the BARS project could benefit all researchers, practitioners, and educators in the community. We also call for generous contributions from the whole community to improve this open benchmarking project and to keep healthy development with the rapid evolution of recommender systems research.

We emphasize that this is not the first and will not be the last work on benchmarking recommender systems. There are a large body of precious research work (e.g.,~\cite{are_we_really,are_we_eval,RecBole,Elliot}, just to name a few) that has been done towards this goal. {In our previous work~\cite{FuxiCTR}, we have built a benchmark specifically for the CTR prediction task. In this work, we initiate the effort of creating an
open benchmarking standard for various recommendation tasks in both retrieval and ranking phases and cover a larger number of models and datasets in the benchmark (detailed in Section~\ref{sec:2_4}).} Yet, implementing and tuning models are very time-consuming and labor-intensive, thus still limiting the scope of the BARS benchmark. We thus would like to call for contributions from the community to build a widely-recognized benchmark for recommender systems.







\begin{figure}[!t]
	\centering
	\includegraphics[width=0.48\textwidth]{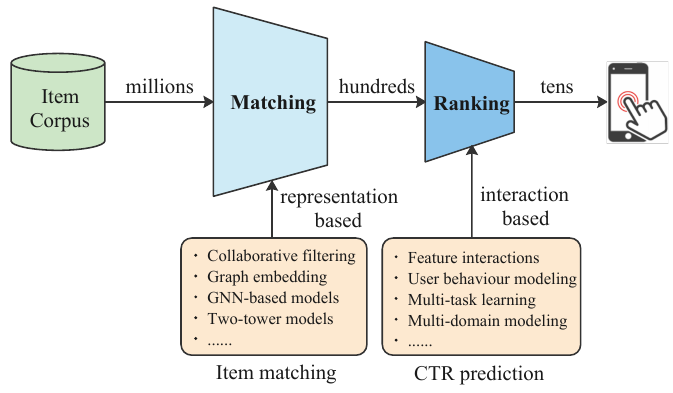}
	\caption{A simplified workflow of modern industrial RS.}
	\label{recsys_arc}
	\vspace{-2ex}
\end{figure}

\section{Background and Related Work}
\subsection{Overview of Recommender Systems}


The goal of recommender systems is to recommend new items to users from a large number of candidate items. Figure~\ref{recsys_arc} illustrates the overall workflow of modern recommender systems. It mainly contains two phases, i.e., retrieval and ranking. The retrieval phase aims to perform candidate item matching from a large item corpus (with millions of items) with high efficiency, while the ranking phase targets at learning the fine-grained personalized ranking of the retrieved items with click-through rate (CTR) prediction\footnote{The ranking phase may also involve other similar types of item scoring, such as CVR prediction, "like" prediction, and reading/watching time prediction} and returning a small number of top-ranking (i.e., in tens) items from the candidates. Through these two phases, the recommender system can strike a balance between accuracy and efficiency. In the following, we introduce the related work of these two recommendation phases.

\subsection{Candidate Item Retrieval}

In the retrieval phase, retrieval efficiency is critically important due to the extremely large item corpus in real systems. Thus, the majority of retrieval algorithms applies representation-based models that enable fast vector retrieval (e.g., via Faiss~\cite{faiss}) and only consider coarse features of users and items. They usually first obtain user and item representations and then compute user-item matching scores through simple similarity measures (e.g., inner product,  cosine). 
The representative methods can be summarized into the several categories, including collaborative filtering (CF)~\cite{CF_survey}, two-tower models~\cite{DSSM,YouTubeNet}, graph embedding models~\cite{deepwalk,LINE}, graph neural networks (GNN)~\cite{PinSage,LightGCN}, and autoencoder-based models~\cite{CDAE,MVAE}.

Collaborative filtering~\cite{CF_survey} methods leverage
the collaborative information among users and items to predict
users’ preferences on candidate items. The classic CF models include neighbourhood-based ones~\cite{ItemKNN} and matrix factorization (MF)-based ones~\cite{MF,GRMF}. 
In recent years, various variants have also been proposed, such as NeuMF~\cite{MF}, ENMF~\cite{ENMF}, and MF-CCL~\cite{SimpleX}. 

The majority of CF methods, however, are focused on learning embeddings of user and item ids. To better incorporate user features and items features in representation learning, two-tower neural networks are often applied. DSSM~\cite{DSSM} is a classic model in which users and items are modelled by two separate network branches. YoutubeDNN~\cite{YouTubeNet} and SimpleX~\cite{SimpleX} can be viewed as simplified variants of two-tower models, as only item ids are used in the item branch. Meanwhile, negative sampling techniques~\cite{reinforced_sampling,mixed_sampling,CBNS} are critical for training such two-tower models. These models ensure high efficiency to fulfill the requirements of practical applications and thus have been widely adopted in industry.

To fully capture the collaborative signals, graph-based algorithms have been widely studied in recent years, including the graph embedding-based models~\cite{deepwalk,ANE,AdvT4NE} and GNN-based models~\cite{PinSage,NGCF,LightGCN}. The reason is that interaction data can be naturally modeled as a user-item bipartite graph, thus candidate item retrieval can be reformulated as a link prediction problem. Graph embedding models (e.g., DeepWalk~\cite{deepwalk},  Node2Vec~\cite{node2vec}, and EGES~\cite{EGES}) usually first learn node representations to capture graph structural information, and then perform link prediction with the learned representations. Currently, GNN-based models, such as PinSage~\cite{PinSage}, LightGCN~\cite{LightGCN}, and UltraGCN~\cite{UltraGCN}, demonstrate quite strong performance for candidate item retrieval due to their effectiveness in modeling graph data. 

Last but not least, autoencoder-based models have also been widely studied for item retrieval, such as CDAE~\cite{CDAE} based on denoising autoencoder, Mult-VAE~\cite{MVAE} based on variational autoencoder, and MacridVAE~\cite{MacridVAE} based on $\beta$-VAE.



While the aforementioned models aim to learn effective representations, how to leverage them for efficiently top-k item retrieval from a large item pool is also a critical problem for candidate item retrieval. For this purpose, many techniques and tools have been developed to support approximate nearest neighbor search (e.g., Faiss~\cite{faiss}, Milvus~\cite{Milvus}, and Scann~\cite{Scann}). 


\begin{table*}[!t]
\centering
\caption{Comparison of existing work (\cmark ~| \textminus ~| \xmark ~means totally | partially | not met, respectively). ``Partially met" indicates incomplete or private availability.}\label{tab:benchmark}
\begin{adjustbox}{max width=1.0\textwidth}
\begin{tabular}{l|c|c|c|c|c|c|c|c|c}
\toprule
Usage & TorchRec & DeepCTR & EasyRec & DeepRec & daisyRec & NeuRec & RecBole  & FuxiCTR & BARS \\ \midrule
Benchmarking tools/datasets & \cmark & \cmark & \cmark  & \cmark & \cmark  & \cmark  & \cmark  & \cmark & \textminus \\
Benchmarking results & \xmark & \xmark & \xmark & \xmark & \textminus & \xmark & \textminus & \xmark & \cmark \\
\bottomrule
\end{tabular}
\end{adjustbox}
\end{table*}

\subsection{CTR Prediction}
The ranking phase aims for personalized ranking of a small set (usually hundreds) of candidate items by considering abundant features of users, items and contexts. CTR prediction is one of the most representative tasks for ranking. 
Due to the practical value of this task, a large number of research efforts have been made in this direction, and many effective models have been designed, ranging from logistic regression models~\cite{LR,FTRL}, to factorization machines~\cite{FM}, to deep neural networks (DNN)~\cite{WideDeep,DeepFM}.

To achieve accurate CTR prediction, feature interaction modeling plays an essential role~\cite{survey_ctr}. A series of DNN-based models have achieved impressive results through designing various feature interaction operators, such as product operators in PNN~\cite{PNN}, NFM~\cite{NFM}, and DCN~\cite{DCN}, convolutional operators in CCPM~\cite{CCPM}, FGCNN~\cite{CNN-FeatureGen}, and FiGNN~\cite{FiGNN}, and attention operators in AFM~\cite{AFM}, FiBiNET~\cite{FiBiNET}, and AutoInt~\cite{AutoInt}. 
Meanwhile, with the success of the wide and deep learning framework, two-stream networks (e.g., Wide\&Deep~\cite{WideDeep}, DeepFM~\cite{DeepFM}, DCN~\cite{DCN}, and AutoInt~\cite{AutoInt}) have been widely applied for CTR prediction tasks, which demonstrate better performance than single-stream networks (e.g., DNN~\cite{YouTubeNet}).

In addition, user behavior modeling serves as an important aspect in CTR prediction. To mine crucial patterns of user interests from historical behaviors, researchers have designed various types of models, ranging from attention-based models~\cite{DIN,DIEN,CIKM_WangZDSZHYB19}, to memory network based-models~\cite{HPMN,UIC}, to retrieval-based models~\cite{UBR,SIM}. More recently, multi-task learning~\cite{ESMM, MMoE} and multi-domain learning~\cite{STAR,One2Many} models have also been widely studied to alleviate the data sparsity issue in CTR prediction. Some other important techniques include large-scale training~\cite{CowClip}, multi-modal fusion~\cite{IMRec}, model ensemble~\cite{EnsembleCTR}, and continual learning~\cite{ReLoop}.

\subsection{Benchmarking and Reproducibility for Recommender Systems}\label{sec:2_4}

Recently, the problems of unreproducible and unfair comparison are raising more and more concerns from the community. 
A series of studies~\cite{are_we_really,are_we_eval,ncf_vs_mf,reenvisioning_ncf_mf} have made in-depth experiments and analysis to point out this serious problem in recommendation, and further call for establishing a unified benchmark.

To encourage more reproducible research, many respectable efforts have been devoted to designing tools or benchmarks. Table~\ref{tab:benchmark} presents a comparison of our BARS project against some representative existing work. We mainly compare them from two perspectives, including: 1) whether the work provides tools or datasets for benchmarking use; 2) whether comprehensive and reproducible benchmarking results are available. It turns out that most of the pioneering efforts focus mainly on building flexible tools for convenient evaluation and benchmarking. Yet, there is still a lack of standard benchmarking pipeline and the comprehensive benchmarking leaderboard.

Among these existing studies, daisyRec~\cite{are_we_eval} highlights the importance of open benchmarking in recommender systems and makes considerable efforts towards this direction by reporting the benchmarking results on some classic recommendation models. Concurrently, RecBole~\cite{RecBole} and Elliot~\cite{Elliot} both present a comprehensive evaluation framework with open-source APIs (e.g., data splitting, data filtering, metrics evaluation, statistical tests, hyperparameter tuning) and RecBole serves tens of SOTA models for reproducible evaluations, which becomes increasingly popular in the community. However, their work fails to provide a benchmark leaderboard of benchmarking results with detailed reproducing steps. 

\begin{figure*}[!t]
	\centering
	\vspace{1ex}
	\includegraphics[width=0.91\textwidth]{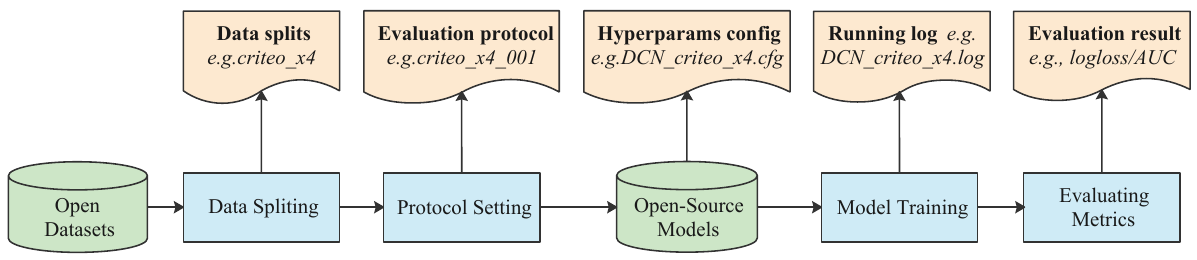}
	\caption{An open benchmarking pipeline for reproducible recommendation research.}
	\label{recsys_pipeline}
\end{figure*}

\underline{\textit{Difference from existing benchmark on CTR prediction~\cite{FuxiCTR}}}: We have established a benchmark for CTR prediction in our previous work~\cite{FuxiCTR}. In comparison, this work aims to make a further step and set up a standard open benchmarking pipeline for a wide range of recommendation tasks in both retrieval and ranking phases. The pipeline and required artifacts for open benchmarking specified in Section~\ref{sec:pipeline} are new to the community and allow researchers to easily expand the benchmarks to more tasks (e.g., re-ranking, sequential recommendation) and vertical scenarios (e.g., news recommendation, music recommendation). They could provide good guidance for researchers and facilitate more rigorous and reproducible research. In detail, we have made the following new contributions. Firstly, our BARS project provides the most comprehensive benchmarking results, w.r.t. coverage of  datasets (10 vs. 4 in~\cite{FuxiCTR}) and models (64 vs. 24 in~\cite{FuxiCTR}) with existing libraries. Specifically, we have integrated more than 70 well-known models covering both retrieval and ranking phases. Secondly, our work proposes a standard open benchmarking pipeline (including 7 key components) as presented in Figure~\ref{recsys_pipeline}, which covers different recommendation tasks. It provides reusable dataset splits, detailed evaluation protocols, convenient open-source models and APIs, complete training logs, and well-tuned benchmarking results for reliable reproducibility of recommendation models. Third, we have built the BARS benchmark website with interactive leaderboards and visualizations for disseminating the benchmarking results. This enables researchers to easily follow and contribute to the leaderboards. 




\section{Open Benchmarking Pipleline}\label{sec:pipeline}
Recommender systems have been an active research area for the past two decades. Unfortunately, there is still a lack of standardized benchmarking and evaluation protocols, which leads to many unreproducbile issues in current studies. Openness in science builds the core for excellence in evidence-based research. Towards the goal for open science, we decide to set up a standardized open benchmarking pipeline for reproducible research in recommender systems. Figure~\ref{recsys_pipeline} presents the open benchmarking pipeline for recommender systems. In summary, the pipeline consists of the following key components.

\textbf{\textit{1) Open datasets}}: Open data are valuable to foster research progress. Fortunately, there have already been a large amount of publicly-available research datasets for recommendation tasks. Yet, most of them are at a small scale, e.g., less than millions of interactions between tens of thousands of users and items. Such data scale differs from the real-world problems in industry. Thus, we encourage researchers to more often choose industrial-scale datasets, which would facilitate to build more practically applicable models and help reduce the gap between research and practice. This is also the goal of this benchmarking work.

\textbf{\textit{2) Data splitting}}: While many possible datasets exist, their uses in the literature are arbitrary, which leads to incomparable and inconsistent results among existing papers. This is largely due to the lack of standardized data splitting and preprocessing. For model evaluation, in many cases, random data splitting is usually performed for training, validation, and testing. Most studies use their own data splits, but no details (neither the random seed nor the preprocessed data) are open, making it difficult for others to reuse the data. Recently, the authors of NGCF~\cite{NGCF}, AutoInt~\cite{AutoInt} and AFN~\cite{AFN} have shared the data processing scripts or the preprocessed data to follow-up researchers, demonstrating good practices for reproducibile research. Our work aims to establish such good practices on data splitting. We assign a unique dataset ID for each specific data split, e.g., Criteo\_x4 denote the unique splits in~\cite{AutoInt}. This could not only facilitate the data reuse in the community according to dataset IDs, but also allow the research results using the same dataset ID directly comparable. 

\textbf{\textit{3) Setting evaluation protocol}}: For each specific data split, we define some specific evaluation protocols to allow for fair comparisons among different models. For example, in Criteo\_x4, we follow the setting in AutoInt~\cite{AutoInt} and set \textit{min\_counts}=10 to filter rare features and feature embedding dimension \textit{d}=16. When a different evaluation protocol (feature embedding \textit{d}=40 as in FGCNN~\cite{CNN-FeatureGen}) is applied, we denote it with a new ID (e.g., Criteo\_x4\_002). Recording such a unique evaluation protocol would allow the reuse of all the benchmarking results for consistent comparisons in future research. 



\textbf{\textit{4) Open-source models}}: In the spirit of open science, many papers have open sourced their model implementations. Even though the official source code may be missing, third-party implementations often exist for an effective model. In many cases, these open-source models could be found at \url{http://paperswithcode.com}. However, despite the availability of model code, the details of hyper-parameter configurations may still be missing or incomplete. This reduce reproducibility because it is not inappropriate to directly run the model with default hyper-parameters. Moreover, most current papers only release their own model code, but not along with their baseline implementations and settings, which largely results in comparisons to weak baselines and thus inconsistent performance increases~\cite{CriticalExam19}. To promote reproducible research, our benchmarking work aims to record detailed hyper-parameter configurations for each experiment and demonstrate the reproducing steps.

\textbf{\textit{5) Model training}}: When training a model, it is beneficial to record the training details into the log, which could help others to quickly glance over the running process of the model. In our work, we advocate to output the running log along with the running environments (e.g., hardware requirements, software versions) as necessary benchmarking artifacts. 

\textbf{\textit{6) Evaluating metrics}}: Evaluation metrics are key to measure model performance and make comparisons. Yet, it is not uncommon that different metrics are used in different papers for evaluation. To standardize the metrics for open benchmarking, we recommend to use Recall@K, HitRate@K, and NDCG@K for candidate item retrieval tasks, and for CTR prediction tasks, we recommend to use logloss and AUC. We defer their definitions in later sections. These metrics are mostly used in the literature and usually adopted in practice. Our benchmarking work reports the results of these standard metrics to allow for easy reuse.

\textbf{\textit{7) Benchmarking artifacts}}: Last but most importantly, our benchmarking pipeline produces the following seven artifacts as illustrated in Figure~\ref{recsys_pipeline}, which are necessary for benchmark reproducibility, including open datasets, dataset ID recording the specific data splitting, the evaluation protocol setting along with the dataset ID, open-source models, hyper-parameter configurations for model training, the running log and evaluation results for each experiment. Lastly, we document the detailed reproducing steps in a friendly format along with these artifacts, in order to ensure that all results are easily reproducible. 

We note that, to the best of our knowledge, this is the first work that aims to build a standard and rigorous pipeline for reproducible research on recommender systems. This imposes new requirements on how researchers deliver their research results and how readers (or reviewers) assess the reproducibility of published papers. Missing any part of the above components defined in the open benchmarking pipeline could largely result in incomplete or broken research artifacts. Unfortunately, most of the existing work under evaluation fails to meet the criteria. We hope that the definition of our open benchmarking pipeline could enhance rigor in current research and offer good guidance for reproducible work.

\section{Benchmarking Candidate Item Retrieval}

\subsection{Benchmarking Settings}

\textbf{Datasets:}
The datasets used for the research of candidate item retrieval are abundant. A lot of work chooses to perform evaluation on their private dataset splits and may use different experimental settings. This makes it difficult to directly make comparisons and accurately measure their progress. 
In this section, we use three representative public datasets, including AmazonBooks~\cite{AmazonBooks_data}, Yelp2018~\cite{Yelp2018}, and Gowalla~\cite{Gowalla}, to present our preliminary benchmarking results.
These three datasets have millions of interactions and are widely adopted to evaluate recommendation models, especially the state-of-the-art GNN-based models (e.g., NGCF~\cite{NGCF}, LightGCN~\cite{LightGCN}). For easy reuse in future research, we uniquely mark the three dataset splits as \texttt{AmazonBooks\_m1}, \texttt{Yelp2018\_m1}, and \texttt{Gowalla\_m1}, respectively. We summarize the statistics of datasets in Table~\ref{tab::matching_dataset}.


\textbf{Evaluation Metrics:} 
We adopt two commonly-used metrics, Recall@k and NDCG@k, in our evaluation. Especially, we set $k=20$ as reported in~\cite{LightGCN}.

\begin{itemize}[leftmargin=5ex]
\item \textbf{Recall}~\cite{Recall_and_Precision}: Recall is the fraction of the user's interested items that are successfully retrieved. Higher recall indicates better retrieval performance for the recommender system.

\item \textbf{NDCG}~\cite{NDCG}: NDCG is a measure of ranking quality. Higher NDCG indicates that more items of users' interests are ranked at top by the recommender system.
\end{itemize}



Note that all the evaluations are performed on the entire item corpus  (i.e., no sampled metrics are used as suggested in~\cite{SampledMetrics}). This also conforms to the practical setting in industry where we usually resort to approximate nearest neighbor search tools (e.g., Faiss~\cite{faiss}) for efficient top-K retrieval even when the entire item corpus is extremely large. The full benchmarking results including more datasets and more metrics are available on the benchmark website.

\textbf{Benchmarked Models:}
We comprehensively evaluate 33 popular models for candidate item retrieval in our benchmark, including 9 collaborative filtering-based models, 4 autoencoder-based models, 3 two-tower based models, 4 graph embedding-based models, and 13 GNN-based models. See Table~\ref{exp:result1} for the models and their references.


\subsection{Benchmarking Results and Analysis}
A systematic performance comparison is shown in Table~\ref{exp:result1}. We show the results of different categories of models in different blocks for clear presentation. The top-5 results under each metric are shown in bold. We obtain several instructive findings from these clear benchmark results.

First, GNN-based item retrieval models are extremely popular in recent years. Performance progress has been gradually made in GNN-based models. From GC-MC, to NGCF, to LightGCN, and to the follow-up models, recommendation has become more and more accurate given more appropriate modeling. Meanwhile, GNN-based models have outperformed many traditional CF models (e.g., NeuMF), which to a certain extent, verifies the potential effectiveness of GNN-based models.

\begin{table}[!t]
\centering
\caption{Datasets statistics for candidate item retrieval.}
\begin{tabular}{c|r|r|r|c}
\toprule
Dataset     & \#Users & \#Items & \#Interactions & Density \\ \midrule
AmazonBooks & 52, 643 & 91, 599 & 2, 984, 108    & 0.062\% \\
Yelp2018    & 31, 668 & 38, 048 & 1, 561, 406    & 0.130\% \\ 
Gowalla     & 29, 858 & 40, 981 & 1, 027, 370    & 0.084\% \\ 
\bottomrule
\end{tabular}
\label{tab::matching_dataset}
\end{table}

\begin{table*}[!t]
\renewcommand\arraystretch{1.068}
\setlength{\tabcolsep}{9pt}
\centering
\caption{Benchmarking results of existing models on candidate item retrieval. We highlight the top-5 best results in each column to indicate the SOTA performance. Note that some rows marked with asterisks are replicated from the existing papers, but all the results are consistently reported under the same evaluation settings.}
\begin{tabular}{rccccccc}
\toprule
\multicolumn{1}{r|}{\multirow{2}{*}{Year}} & \multicolumn{1}{c|}{\multirow{2}{*}{Model}}        & \multicolumn{2}{c|}{AmazonBooks\_m1}         & \multicolumn{2}{c|}{Yelp2018\_m1}            & \multicolumn{2}{c}{Gowalla\_m1} \\ \cline{3-8} 
\multicolumn{1}{r|}{}                             & \multicolumn{1}{c|}{}                              & Recall@20 & \multicolumn{1}{c|}{NDCG@20} & Recall@20 & \multicolumn{1}{c|}{NDCG@20} & Recall@20      & NDCG@20       \\ \hline

\multicolumn{8}{c}{\cellcolor[HTML]{fff8f8}{CF-based Models}}                                                                                                                                                            \\ \hline
\multicolumn{1}{r|}{2001}                     & \multicolumn{1}{c|}{ItemKNN~\cite{ItemKNN}}                       & \hspace{2.5ex}\textbf{0.0736(2)}    & \multicolumn{1}{c|}{\hspace{2.5ex}\textbf{0.0606(1)}}  & 0.0639    & \multicolumn{1}{c|}{0.0531}  & 0.1570         & 0.1214     \\
\multicolumn{1}{r|}{2012}                     & \multicolumn{1}{c|}{MF-BPR~\cite{BPR}}     & 0.0250    & \multicolumn{1}{c|}{0.0196}  & 0.0433    & \multicolumn{1}{c|}{0.0354}  & 0.1291         & 0.1109     \\
\multicolumn{1}{r|}{2017}                    & \multicolumn{1}{c|}{GRMF$^*$~\cite{GRMF}}       & 0.0354    & \multicolumn{1}{c|}{0.0270}  & 0.0571    & \multicolumn{1}{c|}{0.0462}  & 0.1477     & 0.1205     \\
\multicolumn{1}{r|}{2017}                     & \multicolumn{1}{c|}{NeuMF$^*$~\cite{NeuMF}}      & 0.0258    & \multicolumn{1}{c|}{0.0200}  & 0.0451    & \multicolumn{1}{c|}{0.0363}  & 0.1399         & 0.1212     \\
\multicolumn{1}{r|}{2017}                     & \multicolumn{1}{c|}{CML~\cite{CML}}                           & 0.0522    & \multicolumn{1}{c|}{0.0428}  & 0.0622    & \multicolumn{1}{c|}{0.0536}  & 0.1670         & 0.1292     \\
\multicolumn{1}{r|}{2018}                   & \multicolumn{1}{c|}{CMN$^*$~\cite{CMN}}        & 0.0267    & \multicolumn{1}{c|}{0.0218}  & 0.0457    & \multicolumn{1}{c|}{0.0369}  & 0.1405         & 0.1221     \\
\multicolumn{1}{r|}{2018}                  & \multicolumn{1}{c|}{HOP-Rec$^*$~\cite{HOP-Rec}}    & 0.0309    & \multicolumn{1}{c|}{0.0232}  & 0.0517    & \multicolumn{1}{c|}{0.0428}  & 0.1399         & 0.1214     \\ 
\multicolumn{1}{r|}{2020}                    & \multicolumn{1}{c|}{ENMF~\cite{ENMF}}                          & 0.0359    & \multicolumn{1}{c|}{0.0281}  & 0.0624    & \multicolumn{1}{c|}{0.0515}  & 0.1523         & 0.1315     \\ 
\multicolumn{1}{r|}{2021}                    & \multicolumn{1}{c|}{MF-CCL~\cite{SimpleX}}                          & 0.0559    & \multicolumn{1}{c|}{0.0447}  & \hspace{2.5ex}\textbf{0.0698(2)}    & \multicolumn{1}{c|}{\hspace{2.5ex}\textbf{0.0572(2)}}  & \hspace{2.5ex}\textbf{0.1837(5)}         & 0.1493     \\ \hline

\multicolumn{8}{c}{\cellcolor[HTML]{fff8f8}{Autoencoder-based Models}}  \\\hline
\multicolumn{1}{r|}{2011}                    & \multicolumn{1}{c|}{SLIM~\cite{SLIM}}                          & \hspace{2.5ex}\textbf{0.0755(1)}    & \multicolumn{1}{c|}{\hspace{2.5ex}\textbf{0.0602(2)}}  & 0.0646    & \multicolumn{1}{c|}{0.0541}  & 0.1699         & 0.1382     \\
\multicolumn{1}{r|}{2018}                     & \multicolumn{1}{c|}{MultVAE$^*$~\cite{MVAE}}    & 0.0407    & \multicolumn{1}{c|}{0.0315}  & 0.0584    & \multicolumn{1}{c|}{0.0450}  & 0.1641         & 0.1335     \\
\multicolumn{1}{r|}{2019}                 & \multicolumn{1}{c|}{MacridVAE$^*$~\cite{MacridVAE}}  & 0.0383    & \multicolumn{1}{c|}{0.0295}  & 0.0612    & \multicolumn{1}{c|}{0.0495}  & 0.1618         & 0.1202     \\
\multicolumn{1}{r|}{2019}                     & \multicolumn{1}{c|}{EASE\textsuperscript{R}~\cite{EASE}}                       & \hspace{2.5ex}\textbf{0.0710(3)}    & \multicolumn{1}{c|}{\hspace{2.5ex}\textbf{0.0567(4)}}  & 0.0657    & \multicolumn{1}{c|}{0.0552}  & 0.1765         & 0.1467    \\\hline

\multicolumn{8}{c}{\cellcolor[HTML]{fff8f8}{Two-Tower Models}}  \\\hline
\multicolumn{1}{r|}{2016}                  & \multicolumn{1}{c|}{YoutubeDNN~\cite{YouTubeNet}}                    & 0.0502    & \multicolumn{1}{c|}{0.0388}  & \hspace{2.5ex}\textbf{0.0686(4)}    & \multicolumn{1}{c|}{\hspace{2.5ex}\textbf{0.0567(4)}}  & 0.1754         & 0.1473     \\
\multicolumn{1}{r|}{2021}                  & \multicolumn{1}{c|}{SimpleX~\cite{SimpleX}}                    &  0.0583    & \multicolumn{1}{c|}{0.0468}  & \hspace{2.5ex}\textbf{0.0701(1)}    & \multicolumn{1}{c|}{\hspace{2.5ex}\textbf{0.0575(1)}}  & \hspace{2.5ex}\textbf{0.1872(1)}         & \hspace{2.5ex}\textbf{0.1557(3)}    \\ \hline

\multicolumn{8}{c}{\cellcolor[HTML]{fff8f8}{Graph Embedding Models}}                                                                                        \\ \hline
\multicolumn{1}{r|}{2014}                    & \multicolumn{1}{c|}{DeepWalk~\cite{deepwalk}}                      & 0.0346    & \multicolumn{1}{c|}{0.0264}  & 0.0476    & \multicolumn{1}{c|}{0.0378}  & 0.1034         & 0.0740     \\
\multicolumn{1}{r|}{2015}                    & \multicolumn{1}{c|}{LINE~\cite{LINE}}                    & 0.0410    & \multicolumn{1}{c|}{0.0318}  & 0.0549    & \multicolumn{1}{c|}{0.0446}  & 0.1335         & 0.1056     \\
\multicolumn{1}{r|}{2016}                    & \multicolumn{1}{c|}{Node2Vec~\cite{node2vec}}                      & 0.0402    & \multicolumn{1}{c|}{0.0309}  & 0.0452    & \multicolumn{1}{c|}{0.0360}  & 0.1019         & 0.0709     \\
\multicolumn{1}{r|}{2016}                    & \multicolumn{1}{c|}{Item2Vec~\cite{item2vec}}                      & 0.0326    & \multicolumn{1}{c|}{0.0251}  & 0.0503    & \multicolumn{1}{c|}{0.0411}  & 0.1325        & 0.1057    \\\hline

\multicolumn{8}{c}{\cellcolor[HTML]{fff8f8}{GNN-based Models}}                                                                                                                                                                                       \\ \hline
\multicolumn{1}{r|}{2017}                    & \multicolumn{1}{c|}{GC-MC$^*$~\cite{GC-MC}}      & 0.0288    & \multicolumn{1}{c|}{0.0224}  & 0.0462    & \multicolumn{1}{c|}{0.0379}  & 0.1395         & 0.1204     \\
\multicolumn{1}{r|}{2018}                     & \multicolumn{1}{c|}{PinSage$^*$~\cite{PinSage}}    & 0.0282    & \multicolumn{1}{c|}{0.0219}  & 0.0471    & \multicolumn{1}{c|}{0.0393}  & 0.1380         & 0.1196     \\
\multicolumn{1}{r|}{2018}                    & \multicolumn{1}{c|}{GAT$^*$~\cite{GAT}}       & 0.0326    & \multicolumn{1}{c|}{0.0235}  & 0.0543    & \multicolumn{1}{c|}{0.0431}  & 0.1501             & 0.1233          \\
\multicolumn{1}{r|}{2019}                   & \multicolumn{1}{c|}{NGCF$^*$~\cite{NGCF}}       & 0.0344    & \multicolumn{1}{c|}{0.0263}  & 0.0579    & \multicolumn{1}{c|}{0.0477}  & 0.1570         & 0.1327     \\
\multicolumn{1}{r|}{2019}                    & \multicolumn{1}{c|}{DisenGCN$^*$~\cite{DisenGCN}}   & 0.0329    & \multicolumn{1}{c|}{0.0254}  & 0.0558    & \multicolumn{1}{c|}{0.0454}  & 0.1356         & 0.1174     \\
\multicolumn{1}{r|}{2020}                    & \multicolumn{1}{c|}{LR-GCCF~\cite{LR-GCCF}}                       & 0.0335    & \multicolumn{1}{c|}{0.0265}  & 0.0561    & \multicolumn{1}{c|}{0.0343}  & 0.1519         & 0.1285     \\
\multicolumn{1}{r|}{2020}                   & \multicolumn{1}{c|}{NIA-GCN$^*$~\cite{NIA-GCN}}    & 0.0369    & \multicolumn{1}{c|}{0.0287}  & 0.0599    & \multicolumn{1}{c|}{0.0491}  & 0.1359              & 0.1106          \\
\multicolumn{1}{r|}{2020}                   & \multicolumn{1}{c|}{LightGCN$^*$~\cite{LightGCN}}   & 0.0411    & \multicolumn{1}{c|}{0.0315}  & 0.0649    & \multicolumn{1}{c|}{0.0530}  & 0.1830         & \hspace{2.5ex}\textbf{0.1554(4)}     \\
\multicolumn{1}{r|}{2020}                   & \multicolumn{1}{c|}{DGCF$^*$~\cite{DGCF}}       & 0.0422    & \multicolumn{1}{c|}{0.0324}  & 0.0654    & \multicolumn{1}{c|}{0.0534}  & \hspace{2.5ex}\textbf{0.1842(4)}         & \hspace{2.5ex}\textbf{0.1561(2)}    \\
\multicolumn{1}{r|}{2020}                   & \multicolumn{1}{c|}{NGAT4Rec$^*$~\cite{NGAT4Rec}}   & 0.0457    & \multicolumn{1}{c|}{0.0358}  & 0.0675    & \multicolumn{1}{c|}{0.0554}  & --              & --          \\
\multicolumn{1}{r|}{2021}                   & \multicolumn{1}{c|}{SGL-ED$^*$~\cite{SGL-ED}}     & 0.0478    & \multicolumn{1}{c|}{0.0379}  & 0.0675    & \multicolumn{1}{c|}{0.0555}  & --              & --          \\
\multicolumn{1}{r|}{2021}                   & \multicolumn{1}{c|}{GF-CF$^*$~\cite{GF-CF}}     & \hspace{2.5ex}\textbf{0.0710(3)}    & \multicolumn{1}{c|}{\hspace{2.5ex}\textbf{0.0584(3)}}  & \hspace{2.5ex}\textbf{0.0697(3)}    & \multicolumn{1}{c|}{\hspace{2.5ex}\textbf{0.0571(3)}}  & \hspace{2.5ex}\textbf{0.1849(3)}            &  \hspace{2.5ex}\textbf{0.1518(5)}        \\
\multicolumn{1}{r|}{2021}                   & \multicolumn{1}{c|}{UltraGCN~\cite{UltraGCN}}     & \hspace{2.5ex}\textbf{0.0681(5)}    & \multicolumn{1}{c|}{\hspace{2.5ex}\textbf{0.0556(5)}}  & \hspace{2.5ex}\textbf{0.0683(5)}    & \multicolumn{1}{c|}{\hspace{2.5ex}\textbf{0.0561(5)}}  & \hspace{2.5ex}\textbf{0.1862(2)}            &  \hspace{2.5ex}\textbf{0.1580(1)}   \\
\bottomrule
\end{tabular}
\label{exp:result1}
\end{table*}

\begin{table}[!t]
\renewcommand\arraystretch{1.13}
\centering
\caption{Statistics of the CTR prediction datasets.}
\begin{tabular}{c|c|c|c|c}
\toprule
Dataset& \#Instances & \#Fields & \#Features & \%Positives \\
\midrule 
Criteo  & 46M & 39 & 5.55M & 26\% \\
Avazu & 40M & 24 & 8.37M & 17\% \\
KKBox & 6.5M & 13 & 0.1M & 50.4\% \\
\bottomrule
\end{tabular}
\label{tab::ctr_datasets} 
\end{table}

Second, except the graph embedding-based models, the best models in the other four categories are generally comparable. For example, with respect to Recall@20, the autoencoder-based model SLIM achieves the best performance on AmazonBooks while the two-tower based model SimpleX performs best on Yelp2018. Each category has models that perform within top-5, demonstrating that the item retrieval task has not been dominated by any types of methods and still has large development space. New breakthroughs are possible in any category. Furthermore, with the help of this clear benchmark results, researchers may more conveniently identify the bottlenecks and potential improvements of their own models through carefully considering the characteristics of the other models. For example, two well-performed models, i.e., ItemKNN and UltraGCN, leverage the item-item similarity graph, which may be crucial for performance improvement.

Last but most importantly, while many models are proposed with claims of achieving a state-of-the-art performance, they may fail to make a rigorous comparison to some extent. For example, ItemKNN and SLIM, which are two old models that proposed in 2001 and 2011 respectively, can outperform all of the other models on AmazonBooks. Similarly,  YoutubeDNN~\cite{YouTubeNet}, which is a well-known recommendation model proposed in 2016,  surprisingly beats lots of newer models (e.g., HOP-Rec~\cite{HOP-Rec}, MacridVAE~\cite{MacridVAE}, and NIA-GCN~\cite{NIA-GCN}) on the other two benchmark datasets. 
Besides, traditional graph embedding methods, such as DeepWalk, LINE, and Node2Vec, have been less compared, but they have been commonly used in industry as well (e.g.,~\cite{EGES}). In our evaluation, we can see that early GNN-based models, e.g., NGCF, only have comparable (or slightly worse) performance over them. 
In summary, some important baselines are unfortunately omitted by much recent work, resulting in more or less exaggeration of their improvements.  

Such an undesirable phenomenon is largely due to the absence of a unified benchmark. In this case, if the early work accidentally misses some important comparisons, subsequent work based on it may easily repeat this omission that leads to a bad circle.
We envision that our open benchmark will effectively improve this problem and inspire more solid and reproducible work to the community.

\begin{table*}[!t]
\renewcommand\arraystretch{1.03}
\setlength{\tabcolsep}{7.5pt}
\centering
\caption{Benchmarking results of different CTR prediction models. We highlight the top-5 best results in each column. Note that some of the results on Criteo\_x4 and Avazu\_x4 have been reported in our previous work~\cite{FuxiCTR}. Yet, we add substantially more results of new models and new datasets in this paper.}
\begin{tabular}{c|c|cc|cc|cc}
\toprule
 
 &  & \multicolumn{2}{c|}{Criteo\_x4} & \multicolumn{2}{c|}{Avazu\_x4} & \multicolumn{2}{c}{KKBox\_x1} \\ \cline{3-8} 
 
\multirow{-2}{*}{Year} & \multirow{-2}{*}{Model} & Logloss ($\times10^{-2}$) & AUC(\%) & Logloss ($\times10^{-2}$) & AUC(\%) & Logloss ($\times10^{-2}$) & AUC(\%) \\ \midrule
2007 & LR~\cite{manual-2} & 45.68 & 79.34 & 38.15 & 77.75 & 57.46 & 76.78 \\
2010 & FM~\cite{FM} & 44.31 & 80.86 & 37.54 & 78.87 & 50.75 & 82.91 \\
2015 & CCPM~\cite{CCPM} & 44.15 & 81.04 & 37.45 & 78.92 & 50.13 & 83.72 \\
2016 & FFM~\cite{FFM} & 44.07 & 81.13 & 37.20 & 79.31 & 49.74 & 83.76 \\
2016 & HOFM~\cite{HighFM} & 44.11 & 81.07 & 37.54 & 78.91 & 50.48 & 83.15 \\
2016 & PNN~\cite{PNN} & 43.78 & 81.42 & \hspace{2.5ex}\textbf{37.12(4)} & \hspace{2.5ex}\textbf{79.44(3)} & {47.93} & 85.15 \\
2016 & DNN~\cite{YouTubeNet} & 43.80 & 81.40 & 37.22 & 79.28 & 48.11 & 85.01 \\
2016 & Wide\&Deep~\cite{WideDeep} & 43.77 & 81.42 & 37.20 & 79.29 & 48.52 & 85.04 \\
2016 & DeepCrossing~\cite{DeepCross} & 43.84 & 81.35 & 37.21 & 79.30 & 47.99 & 84.95 \\
2017 & NFM~\cite{NFM} & 44.24 & 80.93 & 37.43 & 78.94 & 51.02 & 82.85 \\
2017 & AFM~\cite{AFM} & 44.55 & 80.60 & 37.93 & 78.23 & 52.41 & 81.75 \\
2017 & DeepFM~\cite{DeepFM} & \hspace{2.5ex}\textbf{43.76(3)} & \hspace{2.5ex}\textbf{81.43(5)} & 37.19 & 79.30 & 47.85 & \hspace{2.5ex}\textbf{85.31(4)} \\
2017 & CrossNet~\cite{DCN} & 44.56 & 80.60 & 37.79 & 78.40 & 52.83 & 81.16 \\
2017 & DCN~\cite{DCN} & \hspace{2.5ex}\textbf{43.76(3)} & \hspace{2.5ex}\textbf{81.44(4)} & 37.19 & 79.31 & \hspace{2.5ex}\textbf{47.66(1)} & \hspace{2.5ex}\textbf{85.31(4)} \\
2018 & FwFM~\cite{FwFM} & 44.08 & 81.12 & 37.44 & 79.07 & 49.71 & 84.06 \\
2018 & CIN~\cite{xDeepFM} & 43.94 & 81.27 & 37.42 & 78.94 & 49.09 & 84.26 \\
2018 & xDeepFM~\cite{xDeepFM} & \hspace{2.5ex}\textbf{43.76(3)} & \hspace{2.5ex}\textbf{81.43(5)} & 37.18 & 79.33 & \hspace{2.5ex}\textbf{47.72(2)} & \hspace{2.5ex}\textbf{85.35(2)} \\
2019 & FiGNN~\cite{FiGNN} & 43.83 & 81.38 & 37.36 & 79.15 & 48.96 & 84.72 \\
2019 & FiBiNET~\cite{FiBiNET} & 43.87 & 81.31 & \hspace{2.5ex}\textbf{37.05(2)} & \hspace{2.5ex}\textbf{79.53(2)} & 48.14 & 84.99 \\
2019 & AutoInt~\cite{AutoInt} & 43.99 & 81.19 & 37.45 & 78.91 & 49.19 & 84.36 \\
2019 & AutoInt+~\cite{AutoInt} & 43.90 & 81.32 & 37.46 & 79.02 & \hspace{2.5ex}\textbf{47.73(3)} & \hspace{2.5ex}\textbf{85.34(3)} \\
2019 & HFM~\cite{HFM} & 44.24 & 80.95  & 37.57 & 78.79 & 49.70 & 83.92 \\
2019 & HFM+~\cite{HFM} & 43.92 & 81.27 & 37.14 & \hspace{2.5ex}\textbf{79.44(3)} & \hspace{2.5ex}\textbf{47.81(5)} & {85.21} \\
2019 & FGCNN~\cite{CNN-FeatureGen} & 43.98 & 81.21 & \hspace{2.5ex}\textbf{37.11(3)} & \hspace{2.5ex}\textbf{79.44(3)} & 48.01 & 85.22 \\
2020 & LorentzFM~\cite{LorentzFM} & 44.34 & 80.83 & 37.56 & 78.85 & 51.88 & 82.02 \\
2020 & InterHAt~\cite{InterHAt} & 44.14 & 81.04 & 37.49 & 78.82 & 48.63 & 84.59 \\
2020 & AFN~\cite{AFN} & 44.02 & 81.15 & 37.40 & 79.07 & 49.10 & 84.26 \\
2020 & AFN+~\cite{AFN} & 43.84 & 81.38 & 37.26 & 79.29 & 48.42 & 84.89 \\
2020 & DeepIM~\cite{DeepIM} & \hspace{2.5ex}\textbf{43.75(2)} & \hspace{2.5ex}\textbf{81.46(2)} & \hspace{2.5ex}\textbf{37.16(5)} & 79.35 & \hspace{2.5ex}\textbf{47.75(4)} & \hspace{2.5ex}\textbf{85.37(1)} \\
2020 & ONN~\cite{NFFM} & \hspace{2.5ex}\textbf{43.72(1)} & \hspace{2.5ex}\textbf{81.48(1)} & \hspace{2.5ex}\textbf{36.83(1)} & \hspace{2.5ex}\textbf{79.92(1)} & 48.56 & 84.98 \\
2021 & FmFM~\cite{FmFM} & 43.97 & 81.24 & 37.47 & 79.01 & 49.46 & 84.07 \\
2021 & DCN-V2~\cite{DCN_V2} & \hspace{2.5ex}\textbf{43.75(2)} & \hspace{2.5ex}\textbf{81.45(3)} & {37.19} & 79.31 & 47.87 & \hspace{2.5ex}\textbf{85.31(4)} \\ \bottomrule
\end{tabular}\label{exp:result2}
\vspace{-1.8ex}
\end{table*}

\section{Benchmarking CTR Prediction}
\subsection{Benchmarking Settings}
\textbf{Datasets:} 
CTR prediction has been widely studied in the literature, where models have been evaluated on many different datasets and settings. In this section, we present our preliminary benchmarking results on three widely-used open datasets, including Criteo~\cite{Criteo}, Avazu~\cite{Avazu}, and KKBox~\cite{KKBox}. We summarize the statistics of the datasets in Table~\ref{tab::ctr_datasets}.
In contrast to existing studies that often take their private data splitting and preprocessing, we follow the work~\cite{AutoInt} for random train/validation/test splitting and release the prepossessing code to reduce the bias for benchmarking. The corresponding data splits are uniquely marked as \texttt{Criteo\_x4}, \texttt{Avazu\_x4}, and \texttt{KKBox\_x1}, respectively. We refer readers to our benchmark website for more details about the benchmarking settings and the complete benchmarking results on more datasets.

\textbf{Evaluation Metrics:}
As with many existing studies, we report the results on two most commonly used metrics, i.e., logloss and AUC.

1) \textbf{Logloss}: Logloss is also known as the binary cross-entropy loss, which is used to measure the loss for a binary classification problem. Lower logloss indicates better CTR prediction performance.

2) \textbf{AUC}: AUC~\cite{AUC} is commonly used to measure the probability that a randomly chosen positive sample is ranked higher than a randomly chosen negative sample. Higher AUC means better CTR prediction performance.



\textbf{Benchmarked Models:}
We have comprehensively evaluated a total of 32 models for CTR prediction, which focus primarily on modeling feature interactions. Please refer to Table~\ref{exp:result2} for the list of models and their references.

\subsection{Benchmarking Results and Analysis}
Table~\ref{exp:result2} presents our benchmarking results on three datasets. We also highlight top-5 best results in bold face in each column. We have the following insightful observations from the results. 

First, during the last decade, we have witnessed the evolution of CTR prediction models from simple logistic regression (LR), factorization machines (FM), to various deep models. We can observe that significant performance improvements have been made along this way. Successful deep models include DNN~\cite{YouTubeNet}, Wide\&Deep~\cite{WideDeep}, DeepFM~\cite{DeepFM}, DCN~\cite{DCN}, etc., which have been widely adopted in industry.

Second, we can see that the top-5 best performing models differ widely on different datasets. That is, no one model can rule all the datasets, contradicted with the results reported by existing papers. Nevertheless, we can observe some robustly-performing models including DeepFM~\cite{DeepFM}, DCN~\cite{DCN}, xDeepFM~\cite{xDeepFM}, ONN~\cite{NFFM}, and DCN-V2~\cite{DCN_V2}. We strongly suggest practitioners to try these robust models first for  their practical problems.

Third, we note that recent performance gains of deep models, since the presence of DNN~\cite{YouTubeNet} and Wide\&Deep~\cite{WideDeep}, have become diminished. In many cases, different deep models only make a subtle difference (less than 0.1\%) after sufficient model tuning. We note that the observation is consistent with the experimental results reported in some third-party papers\footnote{As reported in~\cite{DCN_V2}, DNN, PNN, DeepFM, xDeepFM, AutoInt+, and DCN have small AUC differences ($<3e-4$) on Criteo. As reported in~\cite{CowClip}, Wide\&Deep, DeepFM, DCN, and DCN-V2 have similar AUC (differences $<4e-4$) on both Criteo and Avazu.}~\cite{DCN_V2,CowClip}, which validates its correctness. From another perspective, it also reveals that making strong improvements in large-scale CTR prediction tasks is difficult. Therefore, many recent studies tend to improve CTR prediction models from other potential aspects, such as behaviour sequence modeling~\cite{DIEN,DSIN}, multi-task modeling~\cite{ESMM}, and cross-domain modeling~\cite{MiNet,STAR}.


Lastly, we intend to highlight that our benchmarking shows inconsistent results with those reported by some existing papers, which however claim large improvements therein. For example, DeepFM, DCN, xDeepFM, and DCN\_v2 all achieve the same level of accuracy (around 81.4 AUC) on Criteo\_x4, while PNN, HFM+, and FGCNN attain almost the same performance (around 79.44 AUC) on Avazu\_x4. We run many experiments with different hyper-parameters and seeds, but do not obtain sufficiently distinct results. Especially, some recently proposed models, such as InterHAt~\cite{InterHAt}, AFN+~\cite{AFN}, and LorentzFM~\cite{LorentzFM}, obtain even worse results than some previous baselines. This reveals the depressing fact that the performance improvements claimed by some recent papers are somewhat exaggerated. 

Our results raise questions about rigor and reproducibility of the experimental results of existing work, and suggest that baseline models should be more carefully tuned to make more fair comparisons and promote healthy improvements. We hope that the availability of our BARS project could serve as a good starting point and facilitate more reproducible and convincing results in future research.


\section{Call for Contributions}


Openness is key to fostering progress in science via transparency and availability of all research outputs~\cite{openscience}. The evolution of open benchmarking is not possible without active support from the community. Towards this goal, we call for contributions from interested researchers, students, and practitioners. Specifically, everyone can contribute from either one of the following aspects: 1) expanding new datasets, 2) making dataset splits repeatable and shareable, 3) adding new models or new results, 4) verifying the reproducibility of existing models, 5) tuning existing models for better performance, 6) providing valuable feedbacks and suggestions, 7) adopting the open benchmarking pipeline in future research. 

\textbf{Future directions}: While the current benchmark has covered both retrieval and ranking phases of recommender systems, there are still many other tasks that could be incorporated in future versions, such as re-ranking and sequential recommendation. In addition, we could establish specialised benchmarking leaderboards for vertical recommendation scenarios, such as news recommendation, music recommendation, and micro-video recommendation. Moreover, current benchmarking results are only accuracy-focused, it is desirable to add more evaluation metrics such as diversity, coverage, fairness, and so on. Building such a comprehensive open benchmark is a challenging yet worthwhile task for the recommender systems community. We call for active support and contributions in any form from the community to make it possible.

\section{Conclusion}
Reproducibility is an open issue in the field of recommender systems. In this paper, we aim to build an open benchmark for recommender systems, covering both candidate item retrieval and CTR prediction tasks, to drive more solid and reproducible research in this field. Towards this end, we set up a standardized benchmarking pipeline, establish a benchmark website, and deliver the most comprehensive benchmarking results to date. We hope that the project could benefit all researchers, practitioners, and educators in the community.

\begin{acks}
The authors from Tsinghua University are supported in part by the National Natural Science Foundation of China (61972219). We appreciate the support from Mindspore\footnote{\url{https://www.mindspore.cn}}, which is a new deep learning computing framework.
\end{acks}

\balance

\bibliographystyle{ACM-Reference-Format}
\bibliography{bars.bib}
\end{document}